\def\conf{0}
\def\icalp{0}

\ifnum\icalp=0
  \documentclass[11pt]{article}
  \usepackage{fullpage}
\else
  \documentclass{llncs}
\fi

\usepackage{graphicx,amsfonts,amsmath,amssymb,epsfig,hyperref,color}
\usepackage{algorithm}
\usepackage{pdfpages}
\usepackage{enumitem}
\usepackage{multirow}
\usepackage{thmtools}
\usepackage{thm-restate}

\usepackage[noend]{algpseudocode}
\usepackage[font=small,skip=0pt]{caption}

\renewcommand{\paragraph}[1]{{\protect\vspace{8pt}\noindent\sc{#1}}}

\newlength{\saveparindent}
\newlength{\saveparskip}
\newcommand{\BE}{\begin{enumerate}} \newcommand{\EE}{\end{enumerate}}
\newcommand{\BI}{\begin{itemize}} \newcommand{\EI}{\end{itemize}}
\newcommand{\BDes}{\begin{description}}\newcommand{\EDes}{\end{description}}
\newtheorem{alg}{Algorithm}
\newcommand{\BA}{\begin{alg}} \newcommand{\EA}{\end{alg}}
\newtheorem{thm}{Theorem}
\newcommand{\BT}{\begin{thm}} \newcommand{\ET}{\end{thm}}

\newtheorem{lem}{Lemma}      % A counter for Lemmas
\newcommand{\BL}{\begin{lem}} \newcommand{\EL}{\end{lem}}

\newtheorem{fact}{Fact}      % A counter for Lemmas
\newcommand{\BF}{\begin{fact}} \newcommand{\EF}{\end{fact}}
\newtheorem{clm}[lem]{Claim}
\newcommand{\BCM}{\begin{clm}} \newcommand{\ECM}{\end{clm}}

\newtheorem{techcor}[thm]{Corollary}
\newcommand{\BCo}{\begin{techcor}} \newcommand{\ECo}{\end{techcor}}

\newtheorem{cor}[thm]{Corollary}      % counter AS FOR Theorems
\newcommand{\BC}{\begin{cor}} \newcommand{\EC}{\end{cor}}
\newtheorem{prop}[thm]{Proposition}     % A counter AS FOR Thms
\newcommand{\BP}{\begin{prop}} \newcommand {\EP}{\end{prop}}
\newtheorem{conj} {Conjecture}      % counter AS FOR Theorems
\newcommand{\BCJ}{\begin{conj}} \newcommand{\ECJ}{\end{conj}}
\newtheorem{defn}{Definition}         % A counter for Definition
\newcommand{\BD}{\begin{defn}} \newcommand{\ED}{\end{defn}}
\def\FullBox{\hbox{\vrule width 8pt height 8pt depth 0pt}}
\ifnum\icalp=0
\newcommand{\qed}{\;\;\;\FullBox}
\fi
\newcommand{\ourqed}{\;\;\;\FullBox}
\newenvironment{ourproof}{\noindent{\bf Proof:~~}}{\(\ourqed\)}
\newcommand{\BPF}{\begin{ourproof}} \newcommand {\EPF}{\end{ourproof}}
\newenvironment{proofof}[1]{\noindent{\bf Proof of {#1}:~~}}{\(\qed\)}
\newcommand{\BPFOF}{\smallskip \begin{proofof}} \newcommand {\EPFOF}{\end{proofof}}
\newcommand{\qedsketch}{\;\;\;\Box}

\newenvironment{smallproof}{\noindent{\bf Proof:~~}}{\(\qedsketch\)}
\newcommand{\bpf}{\begin{smallproof}} \newcommand{\epf}{\end{smallproof}}
\newcommand{\BEQ}{\begin{equation}} \newcommand{\EEQ}{\end{equation}}
\newcommand{\BEQN}{\begin{eqnarray}}\newcommand{\EEQN}{\end{eqnarray}}
\newcommand{\BEQst}{\begin{equation*}} \newcommand{\EEQst}{\end{equation*}}
\renewcommand{\Pr}{{\rm Pr}}

\newcommand{\poly}{{\rm poly}}

\newcommand{\eps}{\epsilon}

\newcommand{\calR}{{\cal R}}

\newcommand{\calD}{{\cal D}}

\newcommand{\E}{{\rm E}}

\newcommand{\todo}[1]{\iffalse {#1} \fi }

\ifnum\conf=0
\fi

\ifnum\conf=1
\fi
\ifnum\conf=1
\subtitle{Extended abstract}
\else
\fi

\title{Local Computation Algorithms for Graphs of Non-Constant Degrees}
\date{}
\author{
Reut Levi
\thanks{\`{E}cole Normale Sup\`{e}rieure and Universit\`{e} Paris Diderot, France.
E-mail: {\tt  reuti.levi@gmail.com}.
Research supported by NSF grants CCF-1217423 and CCF-1065125, and ISF grants 246/08 and 1536/14.}
\and
Ronitt Rubinfeld
\thanks{CSAIL, MIT, Cambridge MA 02139 and
the Blavatnik School of Computer Science, Tel Aviv University.
E-mail: {\tt  ronitt@csail.mit.edu}.
Research supported by NSF grants CCF-1217423, CCF-1065125, CCF-1420692, and ISF grant 1536/14.}
\and
Anak Yodpinyanee
\thanks{CSAIL, MIT, Cambridge MA 02139.
E-mail: {\tt  anak@csail.mit.edu}.
Research supported by NSF grants CCF-1217423, CCF-1065125, CCF-1420692, and the DPST scholarship, Royal Thai Government.}
}

\begin{document}

\ifnum\icalp=0
\begin{titlepage}
\fi

\maketitle

\ifnum\icalp=0
\thispagestyle{empty}
\fi

\begin{abstract}

In the model of \emph{local computation algorithms} (LCAs), we aim to compute the queried part of the output by examining only a small (sublinear) portion of the input.
Many recently developed LCAs on graph problems achieve time and space complexities with very low dependence on $n$, the number of vertices.
Nonetheless, these complexities are generally at least exponential in $d$, the upper bound on the degree of the input graph.
Instead, we consider the case where parameter $d$ can be moderately dependent on $n$,
and aim for complexities with subexponential dependence on $d$, while maintaining polylogarithmic dependence on $n$.
We present:
\BI
\item a randomized LCA for computing maximal independent sets whose time and space complexities are quasi-polynomial in $d$ and polylogarithmic in $n$;
\item for constant $\eps > 0$, a randomized LCA that provides a $(1-\eps)$-approximation to maximum matching whose time and space complexities are polynomial in $d$ and polylogarithmic in $n$.
\EI

\end{abstract}

\ifnum\icalp=0
\end{titlepage}
\fi

\renewcommand{\baselinestretch}{1.0} 
\setlength{\textfloatsep}{10pt}
\sloppy
\section{Introduction}

In the face of massive data sets, classical algorithmic models, where the algorithm reads the entire input, performs a full computation, then reports the entire output, are rendered infeasible.
To handle these data sets, the model of \emph{local computation algorithms} (LCAs) has been proposed.
As defined in \cite{rubinfeld2011fast}, these algorithms compute the queried part of the output by examining only a small (sublinear) portion of the input.
Let us consider the problem of finding a maximal independent set (MIS) as an example.
The algorithm $\mathcal{A}$ is given access to the input graph $G$, then it is asked a question: ``is vertex $v$ in the MIS?''
The algorithm then explores only a small portion of $G$, and answers ``yes'' or ``no.''
The set of vertices $\{v: \mathcal{A} \textrm{ answers ``yes'' on }v\}$ must indeed form a valid MIS of $G$.
LCAs have been constructed for many problems, including MIS, maximal matching, approximate maximum matching, vertex coloring, and hypergraph coloring
(\cite{rubinfeld2011fast,alon2012space,mansour2012converting,mansour2013local,even2014best,reingold2014new}).
%\mnote{AY: does DR want this fixed or removed?}
%{\color{red}
In our paper, we study MIS and approximate maximum matching; 
these are fundamental graph problems, well-studied in many frameworks, and moreover, 
tools and results for these problems have proven to be useful as building blocks for more sophisticated and specialized problems in the field.
%}

%{\color{red}
The LCA framework is motivated by the circumstances where we focus on computing a small, specified portion of the output.
%}
%\mnote{AY: seems peculiar without the red text}
This key characteristic of LCAs generalizes many other models from various contexts.
For instance, LCAs may take the form of local filters and reconstructors
\cite{chazelle2006online, ailon2008property, brakerski2008local, kale2008noise, saks2010local, jha2013testing, campagna2013local, levi2014local}.
Important applications include locally decodable codes (e.g., \cite{sudan1999pseudorandom}), local decompression \cite{muthukrishnan2005workload, dutta2013simple},
and locally computable decisions for online algorithms and mechanism design \cite{mansour2012converting, hassidim2014local}.
There are a number of works on related models of local computation (e.g., \cite{andersen2006local, borgs2012power, spielman2013local, orecchia2014flow})
as well as lower bounds for the LCA framework and other models (e.g., \cite{gamarnik2014limits}).

Many recently developed LCAs on graph problems achieve time and space complexities with very low dependence on $n$, the number of vertices.
Nonetheless, these complexities are at least exponential in $d$, the upper bound on the degree of the input graph.
While these papers often consider $d$ to be a constant, the large dependence on $d$ may forbid practical uses of these algorithms.
In this work we consider the case where the parameter $d$ can be moderately dependent on $n$,
and provide LCAs for complexities that have quasi-polynomial and even polynomial dependence on $d$, while maintaining polylogarithmic dependence on $n$.
%\mnote{AY: does MM even get mentioned?}
%{\color{blue}
%In particular, Mansour et al.~\cite{mansour2012converting} ask whether there exists an LCA for approximating maximum matching with polynomial dependence on $d$.
%We answer this question in the affirmative. 
%}
%{\color{red}
As noted in \cite{mansour2012converting}, whether there exist LCAs with polynomial dependence on $d$ for these problems is an interesting open question.
Our paper answers this question for the approximate maximum matching problem in the affirmative, and aims at providing techniques useful towards resolving other problems.
%}

\subsection{Related Work}

%{\color{red}
Many useful techniques for designing LCAs originate from algorithms for approximating the solution size in sublinear time.
For example, if we wish to approximate the size of the minimum vertex cover (VC), we may sample a number of vertices and check whether each of them belongs to the minimum VC or not.
The main difference, however, is that an LCA must be able to compute the answer to every query,
while an approximation algorithm is not required to produce a consistent answer for every sample, and may use other properties of the problem to infer its answer.
This paper makes use of a number of common techniques from these approximation algorithms.
%}

In our work we build on the Parnas-Ron reduction, proposed in their paper on approximating the size of a minimum VC \cite{parnas2007approximating}.
This reduction turns a $k$-round distributed algorithm into an LCA by examining all vertices at distance up to $k$ from the queried vertex,
then simulating the computation done by the distributed algorithm, invoking $d^{O(k)}$ queries to the input graph in total.
Using this technique, they obtain a $2$-approximation for the size of a minimum VC (with $\epsilon n$ additive error)
with query complexity $d^{O({\log (d/\eps)})}$, and a $c$-approximation for $c > 2$ using $d^{O(\log d/\eps^3)}$ queries. 
Marko and Ron later improve this result to a $(2+\delta)$-approximation with query complexity $d^{O(\log d)}$ \cite{marko2009approximating}.
Distributed algorithms for the MIS problem require more rounds, and consequently,
a similar reduction only yields an LCA with query and time complexities $d^{O(d \log d)} \log n$ in \cite{rubinfeld2011fast}.
%\mnote{RL: how many rounds?}\mnote{AY: $O(d \log d)$}

%\mnote{AY: no idea what this blue text is saying}
%{\color{blue}
%Many approximation algorithms for graph optimization problems yield an approximation by assuming oracle access to an optimal solution of the optimization problem.
%Given such an oracle, the size of the optimal solution can be approximated through the fraction of randomly sampled vertices that belong to the optimal solution.  
%Usually the oracle is implemented by simulating a localized version of a global algorithm for the problem.
%Each oracle query recursively invokes other oracle queries until the computation is completed. 
%These sequences of recursive queries are referred to as the {\em query trees} of the original queries.   
%By taking this approach, Nguyen and Onak~\cite{nguyen2008constant} simulate the greedy algorithm for maximal matching on the edges of the graph in a random order.
%The answer for each edge only depends on edges preceding it in this order, and thus the number of edges to be investigated can be probabilistically bounded.
%}
%{\color{red}
Another powerful technique for bounding the query and time complexities is the query tree method from the Nguyen-Onak algorithm \cite{nguyen2008constant}.
This method aims to convert \emph{global} algorithms that operate on the entire input into \emph{local} algorithms that adaptively make queries when a new piece of information is needed.
To illustrate this approach, let us consider the MIS problem as an example.
Recall the sequential greedy algorithm where we maintain an initially empty set $I$,
then iterate through the vertex set in some order, adding each vertex to $I$ if it does not have a neighbor in $I$.
Each vertex $v$ will be in the resulting MIS if and only if none of $v$'s neighbors that precede $v$ in our order is already in the MIS.
From this observation, we may create a local simulation of this algorithm as follows: to determine whether $v$ is in the MIS,
we make recursive queries to the preceding neighbors of $v$ and check whether any of them is in the MIS.
As we only query preceding neighbors, the structure of our recursive queries form a \emph{query tree}.
Nguyen and Onak apply this approach on a random order of vertices so that the size of the query tree, which determines the time and query complexities, can be probabilistically bounded.
%}
This method is used in \cite{alon2012space, mansour2012converting, mansour2013local,hassidim2014local, reingold2014new},
giving query complexities with polylogarithmic dependence on $n$ for various problems.
Unfortunately, the expected query tree size is exponential in $d$, which is considered constant in these papers.
For certain problems, a slight modification of the Nguyen-Onak algorithm reduces the expected query tree size to $O(\bar{d})$ \cite{yoshida2009improved, onak2012near}.%\mnote{RL: is this a modification or just improved analysis?} %\mnote{AY: since [NO] is pretty vague about the heuristic, I guess I'd call it a modification ...}
This gives algorithms with $\poly(\bar{d})$ query complexity for approximating the sizes of maximum matching and minimum VC with multiplicative and additive errors.
We build on these results in order to obtain LCAs whose query and time complexities are polynomial in $d$.  

Recently, a new method for bounding the query tree sizes using graph orientation is given in \cite{even2014best} based on graph coloring,
which improves upon LCAs for several graph problems.
They reduce the query complexity of their algorithm for the MIS problem to $d^{O(d^2 \log d)}\log^* n$,
giving the lowest dependence on $n$ currently known, as well as a new direction for developing deterministic LCAs.
%{\color{magenta}
This approach can also be extended back to improve distributed algorithms for certain cases \cite{even2014distributed}.
%}

While all of these LCAs have complexities with exponential dependence on $d$ for the problems studied in this paper, there has been no significant lower bound.
To the best of our knowledge, the only lower bound is of $\Omega(\bar{d})$,
which can be derived from the lower bound for approximation algorithms for the minimum VC problem, given by Parnas and Ron \cite{parnas2007approximating}.

\subsection{Our Contribution and Approaches}

This paper addresses the maximal independent set problem and the approximate maximum matching problem.
The comparison between our results and other approaches is given in table \ref{result-table}.
Our paper provides the first LCAs whose complexities are both quasi-polynomial and polynomial in $d$ and polylogarithmic in $n$ for these problems, respectively.
More concretely, when $d$ is non-constant, previously known LCAs have complexities with polylogarithmic dependence on $n$ only when $d = O(\log \log n)$.
Our LCAs maintain this dependence even when $d = \exp(\Theta((\log \log n)^{1/3}))$ for the MIS problem and $d = \poly(\log n)$ for the approximate maximum matching problem.

\begin{table*}
\centering
\renewcommand{\arraystretch}{1.25} 
\makebox[\textwidth]{%
\begin{tabular}{|p{2.2cm}|c|c|c|c|}
	\hline
	\parbox{2.2cm}{\centering Problem} & Citation & Type & Time & Space \\ \hline
	\multirow{7}{*}{\parbox{2.2cm}{\centering MIS}}
	& \cite{rubinfeld2011fast} 					& randomized 	& $2^{O(d \log^2 d)}\log n$					& $O(n)$ 						\\ \cline{2-5}
	& \cite{alon2012space} 						& randomized 	& $2^{O(d \log^2 d)}\log^3 n$				& $2^{O(d \log^2 d)}\log^2 n$ 	\\ \cline{2-5}
	& \cite{even2014best} 						& deterministic 	& $2^{O(d^2 \log^2 d)} \log^* n^{\dagger}$	& none 							\\ \cline{2-5}
	& \multirow{2}{*}{{\centering \cite{reingold2014new}}} & \multirow{2}{*}{{\centering randomized}}
												 				& $2^{O(d)} \log^2 n$						& $2^{O(d)} \log n \log \log n$ 	\\ \cline{4-5}
	& 											& 				& $2^{O(d)} \log n \log \log n$				& $2^{O(d)} \log^2 n$			\\ \cline{2-5}
%	& reduction on \cite{barenboim2012locality} 	& randomized 	& $2^{O(\log^2 d \cdot \sqrt{\log n})}$		& $O(n \log^2 d)$ 				\\ \cline{2-5}
	& this paper 								& randomized 	& $2^{O(\log^3 d)} \log^3 n$					& $2^{O(\log^3 d)} \log^2 n$ 	\\ \hline
	
%	\multirow{6}{*}{\parbox{2.2cm}{\centering Maximal Matching} }
%	& \cite{mansour2012converting}				& randomized	& $O(\log^3 n)^{\ddagger}$					& $O(\log^3 n)^{\ddagger}$ 		\\ \cline{2-5}
%	& \cite{even2014best} 						& deterministic 	& $2^{O(d^2 \log^2 d)} \log^* n^{\dagger}$	& none 							\\ \cline{2-5}
%	& \multirow{2}{*}{{\centering \cite{reingold2014new}}} & \multirow{2}{*}{{\centering randomized}}
%												 				& $2^{O(d)} \log^2 n$						& $2^{O(d)} \log n \log \log n$ 	\\ \cline{4-5}
%	& 											& 				& $2^{O(d)} \log n \log \log n$				& $2^{O(d)} \log^2 n$			\\ \cline{2-5}
%	& reduction on \cite{barenboim2012locality} 	& randomized 	& $2^{O(\log^2 d + \log d \log^4 \log n)}$		& $O(n \log d)$ 					\\ \cline{2-5}
%	& this paper									& randomized	& $2^{O(\log^2 d)} \log^{3} n$				& $2^{O(\log^2 d)} \log^{2} n$ 	\\ \hline
	
	\multirow{3}{*}{\parbox{2.2cm}{\centering Approximate Maximum Matching}}
	& \cite{mansour2013local} 					& randomized 	& $O(\log^4 n)^{\ddagger}$					& $O(\log^3 n)^{\ddagger}$		\\ \cline{2-5}
	& \cite{even2014best} 						& deterministic 	& $2^{\poly(d)} \poly(\log^* n)^{\dagger}$	& none 							\\ \cline{2-5}
	& this paper 								& randomized 	& $\poly\{d, \log n\}$		& $\poly\{d, \log n\}$ 	\\ \hline
\end{tabular}}

\caption{The summary of complexities of various LCAs. For the approximate maximum matching problem, $\epsilon$ is assumed to be constant.
%Reduction refers to an LCA obtained by directly applying the Parnas-Ron reduction on the distributed algorithm from the cited paper.
$\dagger$ indicates query complexity, when time complexity is not explicitly given in the paper.
$\ddagger$ indicates hidden dependence on $d$, which is at least $2^{O(d)}$ but not explicitly known.}
 \label{result-table}
\end{table*}

%\mnote{RL: I removed the part about exact and approximate algorithms. If our result on MIS implies other results then it would be better to state them.}
%\mnote{AY: I don't understand your comment (2nd sentence).}
%\subsection{Our Approaches}

\subsubsection{Maximal Independent Set}

We provide an LCA for computing a MIS whose time and query complexities are quasi-polynomial in $d$.
We construct a two-phase LCA similar to that of \cite{rubinfeld2011fast}, which is based on Beck's algorithmic approach to Lov\'{a}sz local lemma \cite{beck1991algorithmic}.
In the first phase, we find a large partial solution that breaks the original graph into small connected components.
The LCA for the first phase is obtained by applying the Parnas-Ron reduction on distributed algorithms.
The distributed algorithm for the first phase of the MIS problem in \cite{rubinfeld2011fast} requires $O(d \log d)$ rounds to make such guarantee.
Ours are designed based on recent ideas from \cite{barenboim2012locality} so that $O(\poly(\log d))$ rounds suffice.
As a result, after applying the Parnas-Ron reduction, the query complexity on the first phase is still subexponential in $d$.
Then, in the second phase, we explore each component and solve our problems deterministically; the complexities of this phase are bounded by the component sizes.
By employing a technique from~\cite{alon2012space}, we reduce the amount of space required by our LCA
%This is accomplished through the observation that as we the size of the subgraph explored is small, the required dependence between random bits can be reduced.
%Therefore, we may instead use a construction of $k$-wise independent random variables.
%This reduces the amount of truly random bits from the tape to that of the seed length, which 
so that it has roughly the same asymptotic bound as its time and query complexities.
It is worth mentioning that our LCA for the MIS problem may be extended to handle other problems with reductions to MIS,
such as maximal matching or $(d+1)$-coloring, while maintaining similar asymptotic complexities.

\subsubsection{Approximate Maximum Matching}
We provide an LCA for computing a $(1-\eps)$-approximate maximum matching whose time and query complexities are polynomial in $d$.
Our algorithm locally simulates the global algorithm based on Hopcroft and Karp's lemma \cite{hopcroft1973n}.
This global algorithm begins with an empty matching, then for $\Theta(1/\eps)$ iterations, augments the maintained matching with a maximal set of vertex-disjoint augmenting paths of increasing lengths.
Yoshida et al.~show that this global algorithm has an efficient local simulation in expectation on the queries and the random tapes \cite{yoshida2009improved}.
We derive from their analysis that, on most random tapes, this simulation induces small query trees on most vertices.
From this observation, we construct an efficient two-phase LCA as follows.
In the first phase, we repeatedly check random tapes until we find a good tape such that the query trees for most vertices are small.
This test can be performed by approximating the number of vertices whose query trees are significantly larger than the expected size through random sampling.
In the second phase, we simulate the aforementioned algorithm using the acquired random tape.
On queries for which the query trees are significantly large, we stop the computation and report that those edges do not belong to the matching.
Our good tape from the first phase limits the number of such edges, allowing us to acquire the desired approximation with high probability.

\section{Preliminaries}

\subsection{Graphs}

The input graph $G=(V,E)$ is a simple undirected graph with $|V|=n$ vertices and a bound on the degree $d$, which is allowed to be dependent on $n$.
Both parameters $n$ and $d$ are known to the algorithm.
Let $\bar{d}$ denote the average degree of the graph.
Each vertex $v\in V$ is represented as a unique positive ID from $[n] = \{1,\ldots,n\}$.
For $v \in V$, let $\deg_{G} (v)$ denote the degree of $v$, $\Gamma_G (v)$ denote the set of neighbors of $v$, and $\Gamma^+_G (v) = \Gamma_G (V) \cup \{v\}$.
For $U\subseteq V$, define $\Gamma^+_G (U) = \cup_{u\in U} \Gamma^+_G (u)$. The subscript $G$ may be omitted when it is clear from the context.

We assume that the input graph $G$ is given through an adjacency list oracle $\mathcal{O}^G$ which answers neighbor queries:
given a vertex $v\in V$ and an index $i \in [d]$, the $i^{\rm th}$ neighbor of $v$ is returned if $i \leq \deg (v)$; otherwise, $\bot$ is returned.
For simplicity, we will also allow a degree query which returns $\deg (v)$ when $v$ is given; this can be simulated via a binary-search on $O(\log d)$ neighbor queries.

An \emph{independent set} $I$ is a set of vertices such that no two vertices in $I$ are adjacent.
An independent set $I$ is a \emph{maximal independent set} if no other vertex can be added to $I$ without violating this condition.

A \emph{matching} $M$ is a set of edges such that no two distinct edges in $M$ share a common endpoint.
A matching is a \emph{maximal matching} if no other edge can be added to $M$ out violating this condition.
Let $V(M)$ denote the set of matched vertices, and $|M|$ denote the size of the matching, defined to be the number of edges in $M$.
A \emph{maximum matching} is a matching of maximum size.

\subsection{Local Computation Algorithms}
We adopt the definition of local computation algorithms from \cite{rubinfeld2011fast}, in the context of graph computation problems given an access to the adjacency list oracle $\mathcal{O}^G$.

\BD
A \emph{local computation algorithm} $\mathcal{A}$ for a computation problem is a (randomized) algorithm with the following properties.
$\mathcal{A}$ is given access to the adjacency list oracle $\mathcal{O}^G$ for the input graph $G$, a tape of random bits, and local read-write computation memory.
When given an input (query) $x$, $\mathcal{A}$ must compute an answer for $x$.
This answer must only depend on $x$, $G$, and the random bits.
The answers given by $\mathcal{A}$ to all possible queries must be consistent; namely, all answers must constitute some valid solution to the computation problem.
\ED

The complexities of an LCA $\mathcal{A}$ can be measured in various different aspects, as follows.
\begin{itemize}[noitemsep,nolistsep]
	\item The \emph{query complexity} is the maximum number of queries that $\mathcal{A}$ makes to $\mathcal{O}^G$ in order to compute an answer (to the computation problem) for any single query.
	\item The \emph{time complexity} is the maximum amount of time that $\mathcal{A}$ requires to compute an answer to any single query. We assume that each query to $\mathcal{O}^G$ takes a constant amount of time.
	\item The \emph{space complexity} is total size of the random tape and the local computation memory used by $\mathcal{A}$ over all queries.
	\item The \emph{success probability} is the probability that $\mathcal{A}$ consistently answers all queries.
\end{itemize}

In this paper, we refer to the time and query complexities rather exchangeably: while the time complexity may be much larger than the query complexity in certain cases, for all LCAs considered here, the time complexities are only roughly a factor of $O(\log n)$ larger than the query complexities.
The space complexity of our LCAs are dominated by the size of the random tape, so we often refer to the space complexity as \emph{seed length} instead.
As for the success probability, we consider randomized LCAs that succeed \emph{with high probability}; that is, the success probability can be amplified to reach $1-n^{-c}$ for any positive constant $c$ without asymptotically increasing other complexities.

\subsection{Parnas-Ron Reduction}
Some of our algorithms apply the reduction from distributed algorithms to LCAs proposed by Parnas and Ron \cite{parnas2007approximating}.
This reduction was originally created as a subroutine for approximation algorithms.
Suppose that when a distributed algorithm $\mathcal{A}$ is executed on a graph $G$ with degree bounded by $d$,
each processor (vertex) $v$ is able to compute some function $f$ within $k$ communication rounds under the $\mathcal{LOCAL}$ model.\footnote{In the $\mathcal{LOCAL}$ model, we optimize the number of communication rounds without limiting message size;
in each round, each vertex may send an arbitrarily large message to each of its neighbors.}
Then this function $f$ must only depend on the subgraph of $G$ induced by vertices at distance at most $k$ from $v$.
We can then create an LCA $\mathcal{A'}$ that computes $f$ by simulating $\mathcal{A}$.
Namely, $\mathcal{A'}$ first queries the oracle to learn the structure of this subgraph, then makes the same decision on this subgraph as $\mathcal{A}$ would have done.
In total, this reduction requires $d^{O(k)}$ queries to $\mathcal{O}^G$.

\subsection{Construction of Random Bits and Orderings}

%{\color{magenta}
While our LCAs rely on random bits and orderings, we do not require all bits or orderings to be truly random:
LCAs tolerate some dependence or bias, as they only access a small portion of such random instance in each query.
We now provide some definitions and theorems we will use to construct our LCAs.
%}

\subsubsection{Random Bits}

We will use the following construction to generate $k$-wise independent random bits from the seed (truly random bits) given on the random tape.
\BT[\cite{alon1986fast}]
For $1 \leq k \leq m$, there is a construction of $k$-wise independent random bits $x_1, \ldots, x_m$ with seed length $O(k \log m)$. Furthermore, for $1 \leq i \leq m$, each $x_i$ can be computed in space $O(k \log m)$.
\ET
Note here that each random bit generated from this construction is either 0 or 1 with equal probability.
Nonetheless, for any positive integer $q$, we may generate a random bit that is $1$ with probability exactly $1/q$ using $O(\log q)$ such truly random bits.

\subsubsection{Random Orderings}\label{rand-or-gen}
For $n \geq 1$, let $S_n$ denote the set of all permutations on $[n]$.
Some of our LCAs make use of random permutations of the vertex set.
Generating such uniformly random permutations requires $\Omega(n \log n)$ truly random bits.
Nonetheless, we apply the method from \cite{alon2012space} to construct good random permutations for our LCAs.

We generate our permutations by assigning a random value $r(v)$ to each element $v$, and rank our elements according to these values.
More formally, an \emph{ordering} of $[n]$ is an injective function $r : [n] \rightarrow \calR$ where $\calR$ is some totally ordered set. 
Let $v_1, \ldots, v_n$ be the elements of $[n]$ arranged according to their values mapped by $r$; that is, $r(v_1) < \cdots < r(v_n)$.
We call the permutation $\pi = (v_1, \ldots, v_n)$ of $[n]$ corresponding to this ordering the \emph{projection} of $r$ onto $S_n$.
We may refer to $r(v)$ as the \emph{rank} of $v$.
In our construction, the size of the range of $\calR$ is polynomial in $n$.

A \emph{random ordering} $\calD$ of $[n]$ is a distribution over a family of orderings on $[n]$.
For any integer $2 \leq k \leq n$, we say that a random ordering $\calD$ is $k$\emph{-wise independent}
if for any subset $S \subseteq [n]$ of size $k$, the restriction of the projection onto $S_n$ of $\calD$ over $S$ is uniform over all the $k!$ possible orderings
among the $k$ elements in $S$.
A random ordering $\calD'$ is $\epsilon$\emph{-almost} $k$\emph{-wise independent} if there exists some $k$-wise independent random ordering $\calD$
such that the statistical distance between $\calD$ and $\calD'$ is at most $\epsilon$.

We shall use the following construction from Alon et al.~\cite{alon2012space}.
\BT[\cite{alon2012space}]\label{alg.arv}
Let $n \geq 2$ be an integer and let $2 \leq k \leq n$. Then there is a construction of $(1/n^2)$-almost
$k$-wise independent random ordering over $[n]$ whose seed length is $O(k \log^2 n)$.
\ET
\section{Maximal Independent Set} \label{MISsection}

\subsection{Overview}

Our algorithm consists of two phases.
The first phase of our algorithm computes a large MIS, using a variation of Luby's randomized distributed algorithm \cite{luby1986simple}.
We begin with an initially empty independent set $I$, then repeatedly add more vertices to $I$.
In each round, each vertex $v$ tries to put itself into $I$ with some probability.
It succeeds if none of its neighbors also tries to do the same in that round; 
in this case, $v$ is added to $I$, and $\Gamma^+(v)$ is removed from the graph.

By repeating this process with carefully chosen probabilities, we show that once the first phase terminates,
the remaining graph contains no connected component of size larger than $d^4  \log ⁡n$ with high probability.
This phase is converted into an LCA via the Parnas-Ron reduction.
Lastly, in the second phase, we locally compute an MIS of the remaining graph by simply exploring the component containing the queried vertex.

\subsection{Distributed Algorithm  - Phase 1}

The goal of the first phase is to find an independent set $I$ such that removing $\Gamma^+(I)$ breaks $G$ into components of small sizes.
We design our variation of Luby's algorithm based on Beck's algorithmic approach to Lov{\'a}sz local lemma \cite{beck1991algorithmic};
this approach has been widely applied in many contexts (e.g., \cite{alon1991parallel, kelsen1993fast, rubinfeld2011fast, barenboim2012locality}).
We design our algorithm based on the degree reduction idea from \cite{barenboim2012locality}.\footnote{Applying a similar reduction on the unmodified version gives an LCA with complexities $2^{O(\log^3 d + \log d \log \log n)}$.}
Our algorithm turns out to be very similar to the \textsc{Weak-MIS} algorithm from a recent paper by Chung et al.~\cite{chung2014distributed}.
We state their version, given in Algorithm \ref{dist}, so that we may cite some of their results.

Let us say that a vertex $v$ is \emph{active} if $v \notin \Gamma^+(I)$; otherwise $v$ is \emph{inactive}.
As similarly observed in \cite{peleg2000distributed}, applying a round of Luby's algorithm with selection probability $1/(d+1)$ on a graph with maximum degree at most $d$
makes each vertex of degree at least $d/2$ inactive with constant probability.
To apply this observation, in each iteration, we first construct a graph $G'$ of active vertices.
Next, we apply Luby's algorithm so that each vertex of degree at least $d/2$ becomes inactive with constant probability.
We then remove the remaining high-degree vertices from $G'$ (even if they may still be active).
As the maximum degree of $G$ is halved, we repeat this similar process for $\lceil \log d \rceil$ stages until $G'$ becomes edgeless, where every vertex can be added to $I$.
Since each vertex becomes high-degree with respect to the maximum degree of $G'$ at some stage, each iteration gives every vertex a constant probability to become inactive.

\begin{algorithm}
\caption{Chung et al.'s \textsc{Weak-MIS} algorithm}\label{dist}
\begin{algorithmic}[1]
\Procedure{Weak-MIS}{$G,d$}
	\State $I \gets \emptyset$ \Comment begin with an empty independent set $I$
	\For {iteration $i=1,\ldots,c_1 \log d$} \Comment $c_1$ is a sufficiently large constant
		\State $G' \gets G[V\setminus \Gamma^+(I)]$ \Comment subgraph of $G$ induced by active vertices
		\For {stage $j=1,\ldots,\lceil \log d \rceil$}
			\State $V_j \gets \{v \in V(G') : \deg_{G'}(v) \geq d/2^j\}$ \Comment vertices with degree $\geq$ half of current max.
			\State each $v\in V(G')$ selects itself with probability $p_j = 1/(\frac{d}{2^{j-1}}+1)$ \Comment Luby's algorithm
			\If {$v$ is the only vertex in $\Gamma_{G'}^+ (v)$ that selects itself}
				\State add $v$ to $I$ and remove $\Gamma_{G'}^+ (v)$ from $G'$
			\EndIf
			\State remove $V_j$ from $G'$ \Comment remove high-degree vertices
		\EndFor
		\State add $V(G')$ to $I$ \Comment add lated vertices in $G'$ to $I$
	\EndFor
	\State \textbf{return} $I$
\EndProcedure
\end{algorithmic}
\end{algorithm}

Chung et al.~use \textsc{Weak-MIS} to construct an independent set such that the probability that each vertex remains active is only $1/\textrm{poly}(d)$ \cite{chung2014distributed}.
We cite the following useful lemma that captures the key idea explained earlier. The proof of this lemma is included in Appendix \ref{lemproof}.

\begin{restatable}[\cite{chung2014distributed}]{lem}{revactive}\label{v-active}\label{V-ACTIVE} 
In Algorithm \ref{dist}, if $v \in V_j$, then $v$ remains active after stage $j$ with probability at most $p$ for some constant $p < 1$.
\end{restatable}

Observe that for $v$ to remain active until the end of an iteration, it must be removed in step 10 due to its high degree.
(If $v$ were removed in line 9 or line 11, then $v$ would have become inactive since either $v$ or one of its neighbors is added to $I$.)
Therefore, $v$ must belong to one of the sets $V_j$.
So, each vertex may remain active throughout the iteration with probability at most $p$.
After $\Omega(\log d)$ iterations, the probability that each vertex remains active is only $1/\textrm{poly}(d)$, as desired.

Now we follow the analysis inspired by that of \cite{barenboim2012locality} to prove the guarantee on the maximum size of the remaining active components.
Consider a set $S \subseteq V$ such that $\textrm{dist}_G(u,v) \geq 5$ for every distinct $u,v \in S$.
We say that $S$ is active if every $v\in S$ is active.
As a generalization of the claim above, we show the following lemma.

\begin{restatable}{lem}{resactive}\label{S-active}
Let $S\subseteq V$ be such that $\textrm{dist}_G(u,v) \geq 5$ for every distinct $u,v \in S$, then $S$ remains active until \textsc{Weak-MIS} terminates with probability at most $d^{-\Omega(|S|)}$.
\end{restatable}

\BPF
First let us consider an individual stage. Suppose that $S$ is active at the beginning of this stage.
By Lemma \ref{v-active}, each vertex $v \in S \cap V_j$ remains active after this stage with probability at most $p$.
Notice that for each round of Luby's algorithm, whether $v$ remains active or not only depends on the random choices of vertices within distance $2$ from $v$.
Since the vertices in $S$ are at distance at least $5$ away from one another, the events for all vertices in $S$ are independent.
Thus, the probability that $S$ remains active after this stage is at most $p^{|S\cap V_j|}$.

Now we consider an individual iteration. Suppose that $S$ is active at the beginning of this iteration.
By applying an inductive argument on each stage, the probability that $S$ remains active at the end of this iteration is at most $p^{\sum_{j=1}^{\lceil \log d \rceil} |S\cap V_j|}$.
Recall that for $S$ to remain active after this iteration, every $v\in S$ must belong to some set $V_j$.
So, $\sum_{j=1}^{\lceil \log d \rceil} |S\cap V_j|=|S|$. Thus, $S$ remains active with probability $\exp(-\Omega(|S|))$.

Lastly, we apply the inductive argument on each iteration to obtain the desired bound.
\EPF

Now we are ready to apply Beck's approach to prove the upper bound on the maximum size of the remaining active components (\cite{beck1991algorithmic}, see also \cite{rubinfeld2011fast}).

\BT \label{component-size}
\textsc{Weak-MIS}$(G, d)$ computes an independent set $I$ of an input graph $G$ within $O(\log^2 d)$ communication rounds, such that the subgraph of $G$ induced by active vertices contains no connected component of size larger than $d^4  \log ⁡n$ with probability at least $1-1/\poly(n)$.
\ET
%{\color{magenta}
\BPF
We provide a proof sketch of this approach.
Let $T$ be a tree embedded on the \emph{distance-}5 graph, defined as $(V, \{(u, v): \textrm{dist}_G(u,v) = 5\})$, such that $\textrm{dist}_G(u,v) \geq 5$ for every distinct $u,v \in V(T)$. 
By Lemma \ref{S-active}, the probability that its vertex set $S$ remains active is $d^{-\Omega(s)}$, where $s = |S|$.
It can be shown combinatorially that there are at most $n(4d^5)^s$ distinct trees of size $s$ embedded on the distance-5 graph (for more details, see the proof of Lemma 4.6 in~\cite{rubinfeld2011fast}). 
Therefore, the expected number of such trees whose vertex sets remain active is at most $n(4d^5)^s \cdot d^{-\Omega(s)}$.
For $s = \log n$, this quantity is bounded above by $1/\textrm{poly}(n)$.
By Markov's inequality, all such sets $S$ are inactive with probability at least $1-1/\poly(n)$.

Observe that any connected subgraph of $G$ of size at least $d^4 \log n$ must contain some set $S$ satisfying the aforementioned condition.
Thus, the probability that any such large set remains active after \textsc{Weak-MIS} terminates is also bounded above by $1/\textrm{poly}(n)$.
\EPF
%}

\subsection{Constructing the LCA - Phase 1}

We now provide the Parnas-Ron reduction of \textsc{Weak-MIS} in Algorithm \ref{dist} into an LCA.
Let us start with a single stage, given as procedure \textsc{LC-MIS-Stage} in Algorithm \ref{lca-stage}.
The given parameters are the graph $G'$ (via oracle access), the degree bound $d$, the queried vertex $v$, the iteration number $i$ and the stage number $j$.
Given a vertex $v$, this procedure returns one of the three states of $v$ at the end of iteration $i$, stage $j$:
\begin{itemize}[noitemsep,nolistsep]
	\item YES if $v\in I$ (so it is removed from $G'$)
	\item NO if $v\notin I$ and it is removed from $G'$
	\item $\bot$ otherwise, indicating that $v$ is still in $G'$
\end{itemize}

\begin{algorithm}
\caption{LCA of Phase 1 for a single stage of \textsc{Weak-MIS}}\label{lca-stage}
\begin{algorithmic}[1]
\Procedure{LC-MIS-Stage}{$\mathcal{O}^{G'},d,v,i,j$}
	\State \textbf{if} {$j = 0$} \textbf{then} \textbf{return} $\bot$ \Comment{every vertex is initially in $G'$}
	\If {\textsc{LC-MIS-Stage}$(\mathcal{O}^{G'},d,v,i,j-1) \neq \bot$}
		\State \textbf{return} \textsc{LC-MIS-Stage}$(\mathcal{O}^{G'},d,v,i,j-1)$  \Comment{$v$ is already removed from $G'$}
	\EndIf
	\For {each $u$ within distance 2 from $v$ on $G'$} \Comment{check the status of every vertex near $v$}
		\If {\textsc{LC-MIS-Stage}$(\mathcal{O}^{G'},d,u,i,j-1) = \bot$}
			\State status$(u) \gets$ removed
		\Else
			\State \textbf{if} {$B(u,i,j) = 1$} \textbf{then} status$(u) \gets$ selected \Comment{$B(u,i,j)=1$ means $u$ chooses itself}
			\State \textbf{else} status$(u) \gets$ not selected
		\EndIf
	\EndFor
	\If {status$(v) = $ selected AND $\forall u\in\Gamma_{G'}(v)$: status$(u)\neq$ selected}
		\State \textbf{return} YES \Comment {$v$ is added to $I$}
	\EndIf
	\If {status$(v) = $ not selected}
		\If {$\exists u\in\Gamma_{G'}(v)$: status$(u) = $ selected AND $\forall w\in\Gamma_{G'}(u)$: status$(w)\neq$ selected}
			\State \textbf{return} NO \Comment {a neighbor $u$ of $v$ is added to $I$, so $v$ is removed}
		\EndIf
	\EndIf
	\If {$|\{u\in\Gamma_{G'}(v): \textrm{status}(u) \neq$ removed$\}| \geq d/2^j$}
		\State \textbf{return} NO \Comment {$v$ is removed due to its high degree}
	\EndIf
	\State \textbf{return} $\bot$ \Comment {$v$ remains in $G'$}
\EndProcedure
\end{algorithmic}
\end{algorithm}

For simplicity, we assume that the local algorithm has access to random bits in the form of a publicly accessible function
$B: V\times[c_1 \log ⁡d] \times [\lceil \log d \rceil]\rightarrow\{0,1\}$ such that $B(v,i,j)$ returns $1$ with the selection probability $p_j$ (from Algorithm \ref{dist}, line 7) and returns $0$ otherwise.
This function can be replaced by a constant number of memory accesses to the random tape,
and we will explore how to reduce the required amount of random bits in Section \ref{reduce-mem}.
Note also that we are explicitly giving the oracle $\mathcal{O}^{G'}$ as a parameter rather than the actual graph $G'$.
This is because we will eventually simulate oracles for other graphs, but we never concretely create such graphs during the entire computation.
Observe that each call to \textsc{LC-MIS-Stage} on stage $j$ invokes up to $O(d^2)$ calls on stage $j-1$.
Simulating all stages by invoking this function with $j = \lceil \log d \rceil$ translates to $d^{O(\log d)}$ queries to the the base level $j=0$.

Next we give the LCA \textsc{LC-MIS-Iteration} for computing a single iteration in Algorithm \ref{lca-iteration}.
The parameters are similar to that of \textsc{LC-MIS-Stage}, but the return values are slightly different:
\begin{itemize}[noitemsep,nolistsep]
	\item YES if $v\in I$ (so it is inactive)
	\item NO if $v\in \Gamma^+_G(I)$ (so it is inactive)
	\item $\bot$ otherwise, indicating that $v$ is still active
\end{itemize}
In case $v$ is still active by the end of iteration $i-1$, we must simulate iteration $i$ using \textsc{LC-MIS-Stage}.
We must return YES if $v \in I$ (line 7 of Algorithm \ref{lca-iteration}); which may occur in two cases.
It can be added to $I$ in some stage, making \textsc{LC-MIS-Stage} returns YES (corresponding to line 9 of Algorithm \ref{dist}).
It may also remain in $G'$ through all iterations, making \textsc{LC-MIS-Stage} return $\bot$; $v$ is then added to $I$ because it is isolated in $G'$ (line 11 of Algorithm \ref{dist}).

\begin{algorithm}
\caption{LCA of Phase 1 for a single iteration of \textsc{Weak-MIS}}\label{lca-iteration}
\begin{algorithmic}[1]
\Procedure{LC-MIS-Iteration}{$\mathcal{O}^{G},d,v,i$}
	\State \textbf{if} {$i = 0$} \textbf{then} \textbf{return} $\bot$ \Comment{$I$ is initially empty}
	\If {\textsc{LC-MIS-Iteration}$(\mathcal{O}^G,d,v,i-1) \neq\bot$}
		\State \textbf{return} \textsc{LC-MIS-Iteration}$(\mathcal{O}^G,d,v,i-1)$ \Comment{$v$ is already inactive}
	\EndIf
	\State $\mathcal{O}^{G'} \gets $ oracle for the subgraph induced by $\{u : $ \textsc{LC-MIS-Iteration}$(\mathcal{O}^G,d,u,i-1) = \bot\}$
	\If {\textsc{LC-MIS-Stage}$(\mathcal{O}^{G'},d,v,i,\lceil \log d \rceil) \neq $ NO}
		\State \textbf{return} YES \Comment{$v$ is added to $I$ in some stage (YES) or at the end ($\bot$)}
	\ElsIf {$\exists u\in\Gamma_{G}(v)$: \textsc{LC-MIS-Stage}$(\mathcal{O}^{G'},d,u,i,\lceil \log d \rceil) \neq $ NO}
		\State \textbf{return} NO \Comment{a neighbor of $u$ of $v$ is added to $I$}
	\EndIf
	\State \textbf{return} $\bot$ \Comment{$v$ is still active}
\EndProcedure
\end{algorithmic}
\end{algorithm}

To implement this LCA, we must simulate an adjacency list oracle for $G'$ using the given oracle for $G$, which can be done as follows.
For a query on vertex $v$, we call \textsc{LC-MIS-Iteration}$(\mathcal{O}^G,d,u,i-1)$ on all vertices $u$ at distance at most 2 away from $v$.
This allows us to determine whether each neighbor of $v$ is still active at the beginning of iteration $i$, as well as providing degree queries.
We then modify the ordering of the remaining neighbors (by preserving the original ordering, for example) to consistently answer neighbor queries.
That is, $\mathcal{O}^{G'}$ can be simulated through at most $O(d^2)$ function calls of the form \textsc{LC-MIS-Iteration}$(\mathcal{O}^G,d,u,i-1)$.

Using this subroutine, the LCA \textsc{LC-MIS-Phase1} for \textsc{Weak-MIS} can be written compactly as given in Algorithm \ref{lca-phase1}.
Via a similar inductive argument, we can show that each call to \textsc{LC-MIS-Phase1} translates to $d^{O(\log^2 d)} = 2^{O(\log^3 d)}$ calls to the original oracle $\mathcal{O}^{G}$.
The running time for the LCA is clearly given by the same bound.
The memory usage is given by the amount of random bits of $B$, which is $O(n \log^2 d)$.
We summarize the behavior of this LCA through the following lemma.

\BL \label{phase1-lemma}
\textsc{LC-MIS-Phase1} is a local computation algorithm that computes the distributed \textsc{Weak-MIS} algorithm with time complexity $2^{O(\log^3 d)}$ and space complexity $O(n \log^2 d)$.
\EL

\begin{algorithm}
\caption{LCA of Phase 1 (\textsc{Weak-MIS})}\label{lca-phase1}
\begin{algorithmic}[1]\Procedure{LC-MIS-Phase1}{$\mathcal{O}^{G},d,v$}
	\State \textbf{return} \textsc{LC-MIS-Iteration}$(\mathcal{O}^G,d,v,c_1 \log d)$
\EndProcedure
\end{algorithmic}
\end{algorithm}

\subsection {Constructing the LCA - Phase 2 and the Full LCA}

Let $G''$ be graph $G$ induced by active vertices after \textsc{Weak-MIS} terminates.
By Theorem \ref{component-size}, with high probability, $G''$ contains no component of size exceeding $d^4 \log n$.
Therefore, to determine whether an active vertex $v$ is in the MIS,
we first apply a breadth-first search until all vertices in $C(v)$, the component containing $v$, are reached.
Then we compute an MIS deterministically and consistently (by choosing the lexicographically first MIS, for instance).
This procedure is summarized as $\textsc{LC-Phase2}$ in Algorithm \ref{lca-phase2}.

\begin{algorithm}
\caption{LCA of Phase 2}\label{lca-phase2}
\begin{algorithmic}[1]\Procedure{LC-Phase2}{$\mathcal{O}^{G''},d,v$}
	\State breadth-first search for $d^4 \log n$ steps on $G''$ to find $C_v$
	\State \textbf{if} {$|C_v| > d^4 \log n$} \textbf{then} \textbf{report} ERROR \Comment only occurs with probability $1/\poly(n)$
	\State deterministically compute an MIS $I_{C_v}$ of $C_v$
	\State \textbf{if} {$v \in I_{C_v}$} \textbf{then} \textbf{return} YES
	\State \textbf{else} \textbf{return} NO
\EndProcedure
\end{algorithmic}
\end{algorithm}

The algorithm only reports ERROR when the component size exceeds the bound from Theorem \ref{component-size}.
Clearly, a call to $\textsc{LC-Phase2}$ makes at most $\poly(d) \log ⁡n$ queries to $\mathcal{O}^{G''}$ in total.
The lexicographically first maximal independent set of $C_v$ can be computed via a simple greedy algorithm with $O(d\cdot|C_v|)$ time complexity.
Overall, both time and query complexities are $\textrm{poly}\{d,\log n\}$.

Combining both phases, we obtain the LCA \textsc{LC-MIS} for computing an MIS as given in Algorithm \ref{lca-mis}.
We now prove the following theorem.

\begin{algorithm}
\caption{LCA for computing a maximal independent set}\label{lca-mis}
\begin{algorithmic}[1]\Procedure{LC-MIS}{$\mathcal{O}^{G},d,v$}
	\If {\textsc{LC-MIS-Phase1}$(\mathcal{O}^G,d,v) \neq\bot$}
		\State \textbf{return} \textsc{LC-MIS-Phase1}$(\mathcal{O}^G,d,v)$ \Comment{$v$ is already inactive}
	\EndIf
	\State $\mathcal{O}^{G''} \gets $ oracle for the subgraph induced by $\{u : $ \textsc{LC-MIS-Phase1}$(\mathcal{O}^G,d,u) = \bot\}$
	\State \textbf{return} \textsc{LC-MIS-Phase2}$(\mathcal{O}^{G''},d,v)$
\EndProcedure
\end{algorithmic}
\end{algorithm}

\BT \label{thm-mis}
There exists a randomized local computation algorithm that computes a maximal independent set of $G$ with time complexity $2^{O(\log^3 d)} \log n$ and space complexity $O(n \log^2 d)$.
\ET
\BPF
To obtain the time complexity, recall from Lemma \ref{phase1-lemma} that each call to \textsc{LC-Phase1} can be answered
within $2^{O(\log^3 d)}$ time using $2^{O(\log^3 d)}$ queries to $\mathcal{O}^G$.
Thus the adjacency list oracle $\mathcal{O}^{G''}$ can be simulated with the same complexities. 
The LCA for Phase 2 makes $\textrm{poly}(d) \log n$ queries to $\mathcal{O}^{G''}$, resulting in $2^{O(\log^3 d)} \log n$ total computation time and queries.
The required amount of space is dominated by the random bits used in Phase 1.
The algorithm only fails when it finds a component of size larger than $d^4 \log n$, which may only occur with probability $1/\poly(n)$ as guaranteed by Theorem \ref{component-size}.
\EPF

\subsection{Reducing Space Usage} \label{reduce-mem}

In this section, we directly apply the approach from \cite{alon2012space} to reduce the amount of random bits used by our LCA, thus proving the following theorem.

\begin{restatable}{thm}{resmem}\label{thm-mem}
There exists a randomized local computation algorithm that computes a maximal independent set of $G$ with ra seed of length $2^{O(\log^3 d)} \log^2 n$ and time complexity $2^{O(\log^3 d)} \log^3 n$.
\end{restatable}

\BPF
Observe that the MIS algorithm in Theorem \ref{thm-mis} constructed throughout this section does not require fully independent random bits in function $B$.
Our algorithm can answer a query for any vertex $v$ by exploring up to $q = 2^{O(\log^3 d)} \log n$ vertices in total.
So, random bits used by vertices not explored by a query do not affect our answer.
One bit is used by each vertex in each round of Phase 1, and thus we only create $b = O(\log^2 d)$ random bits for each vertex.
Therefore, out of $O(n \log^2 d)$ bits from function $B$, only $q\cdot b$ bits are relevant for each query.

To generate random bits for function $B$, we will apply Theorem \ref{rand-or-gen}.
As we may require bits of function $B$ to be 1 with probability as low as $1/(d+1)$,
we will need up to $\lceil\log d \rceil$ bits from this construction to obtain one bit for function $B$.
So in our case, we have $k = \lceil\log d \rceil \cdot q\cdot b = 2^{O(\log^3 d)} \log n$ and $m = \lceil\log d \rceil \cdot O(n \log^2 d) = O(n \log^3 d)$.
Thus the amount of space can be reduced to $O(k \log m) = 2^{O(\log^3 d)} \log n \cdot \log(n \log^3 d) = 2^{O(\log^3 d)} \log^2 n$.
Similarly, the required amount of time for computing each random bit becomes $2^{O(\log^3 d)} \log^2 n$.
\EPF

\subsection{Other Remarks} \label{mmsec}

The proposed LCA for the maximal independent set problem can be used as a building block for other problems.
For example, a $(d+1)$-coloring can be computed by finding a maximal independent set of $G \times K_{d+1}$ where $K_{d+1}$ is a clique of size $d+1$, resulting in an LCA within the same asymptotic complexities (see also \cite{linial1992locality}).

We may apply our MIS algorithm to the line graph $L(G)$ in order to compute a maximal matching on $G$ within the same asymptotic complexities as Theorem \ref{thm-mem}.
Nonetheless, Barenboim et al.~provide a distributed routine for computing a large matching that yields a similar bound as Theorem \ref{component-size}
that requires only $O(\log d)$ rounds \cite{barenboim2012locality}.
We apply the Parnas-Ron reduction on their algorithm as similarly done in Section \ref{MISsection} to obtain the following theorem,
where the dependence on $d$ is reduced from $2^{O(\log^3 d)}$ to $2^{O(\log^2 d)}$.
Additional details for this result can be found in Appendix \ref{thm-mm2-proof}.

\begin{restatable}{thm}{resmmt}\label{thm-mm2}\label{THM-MM2}
There exists a randomized local computation algorithm that computes a maximal matching of $G$ with seed of length $2^{O(\log^2 d)}\log^{2} n$ and time complexity $2^{O(\log^2 d)}\log^{3} n$.
\end{restatable}
\section{Approximate Maximum Matching}
\subsection{Overview}
In this section, we aim to construct an LCA that provides a $(1-\epsilon)$-approximation to maximum matching.
To do so, we first address the simpler problem of computing a $(1-\epsilon)$-approximation to MIS.
We present this LCA because it allows us to explain our LCA for the approximate maximum matching problem in a much clearer manner,
as they share both their main principles and the two-phase structure.
Moreover, this LCA is useful as a subroutine for other problems.
For example, an approximation to MIS of the line graph $L(G)$ readily yields a $(1/2-\eps)$-approximation to maximum matching of $G$.

\subsection{LCA for Computing an Approximate Maximal Independent Set}
We now describe an LCA that provides a $(1-\eps)$-approximation to MIS. 
More specifically, with high probability, the set of vertices $\tilde{I}$ that our LCA answers YES must be a subset of some maximal independent set $I$ satisfying $|\tilde{I}| \geq (1-\eps)|I|$.
Our algorithm is based on a local simulation of the classical greedy algorithm for computing an MIS.
This greedy algorithm iterates over the vertex set according to some arbitrary order and adds a vertex to the
constructed independent set if and only if none of its neighbors has been added previously.

Algorithm~\ref{alg.MIS} summarizes the local simulation of the greedy algorithm as suggested by Nguyen and Onak and further analyzed by Yoshida et al.~\cite{nguyen2008constant,yoshida2009improved}. 
In addition to the adjacency list oracle $\mathcal{O}^G$ and the queried vertex $v$,
$\textsc{LS-MIS}$ also receives as an input a permutation $\pi$ on the vertices.
The set of vertices $v$ such that $\textsc{LS-MIS}(\mathcal{O}^G, \pi, v)$ returns YES is the lexicographically first MIS according to $\pi$,
which is also the MIS outputted by the greedy algorithm when the vertices are iterated in this order.

\begin{algorithm}
\caption{Local simulation of the greedy MIS algorithm}
\label{alg.MIS}
\begin{algorithmic}[1]
\Procedure{LS-MIS}{$\mathcal{O}^G, \pi, v$}
\State query $\mathcal{O}^G$ for all neighbors of $v$
\State let $v_1, \ldots, v_k$ be the neighbors of $v$ sorted according to $\pi$
\For  {$i=1,\ldots, k$}
\If{$v_i$ precedes $v$ in $\pi$} 
	\State compute $\textsc{LS-MIS}(\mathcal{O}^G, \pi, v_i)$
\If{$\textsc{LS-MIS}(\mathcal{O}^G, \pi, v_i) =$ YES} \textbf{return} NO \EndIf  \EndIf 
\EndFor
\State \textbf{return} YES 
\EndProcedure
\end{algorithmic}
\end{algorithm}

We now give an overview of the construction of our LCA.
From the analysis of Yoshida et al.~we obtain that in expectation over the queries and the random permutations, the query complexity of Algorithm~\ref{alg.MIS} is polynomial in $d$, the degree bound \cite{yoshida2009improved}.
More formally, let $R^G_{\pi}(v)$ denote the number of (recursive) calls to $\textsc{LS-MIS}$ during the evaluation of $\textsc{LS-MIS}(\mathcal{O}^G, \pi, v)$.
This result from Yoshida et al.~can be succinctly stated as follows.

\BT[\cite{yoshida2009improved}] \label{YYI}
For any graph $G = (V, E)$ with $n$ vertices and $m$ edges, 
\BEQst
\E_{\pi \in S_n, v \in V} [R^G_{\pi}(v)] \leq 1 + \frac{m}{n}\;. \label{eq.1}
\EEQst
\ET

However, the query complexity of an LCA is determined by the maximum number of queries invoked in order to answer any single query, so this expected bound does not readily imply an efficient LCA.
We may alter Algorithm~\ref{alg.MIS} so that it computes an independent set that is relatively large but not necessarily maximal as follows.
If a query on a vertex $v$ recursively invokes too many other queries, then the simulation is terminated and the algorithm immediately returns NO, i.e., that $v$ does not belong to our independent set.
We show that Theorem \ref{YYI} implies that for at least a constant fraction of the orderings, this modification yields a good approximation ratio. 
Nonetheless, an LCA must succeed with high probability rather than only a constant probability.
We resolve this issue by repeatedly sampling permutations until we obtain a sufficiently good one.

While generating a truly random permutation $\pi$ requires $\Omega(n)$ random bits,
we show that we may generate sufficiently good permutations by projecting random orderings from a distribution of small entropy onto $S_n$. 
Using the construction from Alon et al.~as stated in Theorem \ref{alg.arv}, it follows that a small seed whose size is polylogarithmic in $n$ suffices for our algorithm.
From these outlined ideas, we are now ready to prove the following result.

\BT\label{thm.mis}
There exists a randomized $(1-\eps)$-approximation local computation algorithm for maximal independent set with random seed of length $O((d^2/\eps^2)\log^2 n \log \log n)$ and query complexity $O((d^4/\eps^2) \log^2 n \log \log n)$.
\ET

\BPF
Our algorithm builds on Algorithm~\ref{alg.MIS} and consists of two phases as follows.
The input of the algorithm is a random seed (tape) $s$ and a queried vertex $v\in V$.
In the first phase, using $s$ and graph queries, we find a good permutation $\pi = \pi_G(s)$ over the vertex set.
Note that this good permutation $\pi$ is dependent only on $s$ and $G$, and therefore must be the identical for any query. 
Then in the second phase, we decide whether $v$ belongs to our independent set using the altered local simulation that limits the number of invoked queries.
 
Formally, we say that $\pi \in S_n$ is {\em good} if $\Pr_{v\in V}\left[ R^G_{\pi}(v) > \ell\right] \leq \gamma$ where $\gamma = \eps/d$, $\ell = 6t/\eps$ and $t = m/n + 1$.
Let $\delta$ denote the error probability. %(For our purposes, we will later substitute $\delta = 1/\poly(n)$ to obtain an LCA that succeeds with high probability.)
We claim that, given access to $D_\ell$, an $\ell$-wise independent ordering constructed from $s$, Algorithm \ref{alg.MIS.1} computes a good permutation $\pi$ with the desired success probability.
This algorithm checks each constructed permutation $\pi_i$ whether it is good by approximating the fraction $p_{\pi_i}$ of vertices whose induced query trees have size exceeding $\ell$.
The set $S$ of sampled vertices is sufficiently large that we obtain a good approximation $\tilde{p}_{\pi_i}$ of $p_{\pi_i}$,
and note that checking whether $R^G_{\pi_i}(v) > \ell$ can be accomplished by simply simulating Algorithm \ref{alg.MIS} until $\ell$ recursive calls are invoked.

\begin{algorithm}
\caption{Phase 1 of the LCA for the approximate MIS algorithm (finding a good ordering)}
\label{alg.MIS.1}
\begin{algorithmic}[1]
\Procedure{LC-AMIS-Phase1}{$\mathcal{O}^G, s$}
\State let $S$ be a multi-set of $\Theta(\log (1/\delta)\log \log (1/\delta)/\gamma^2)$ vertices chosen uniformly at random
\For  {$i=1,\ldots, \Theta(\log (1/\delta))$}
	\State let $\pi_i$ be the projection of an ordering independently drawn from $D_\ell$
	\State let $\tilde{p}_{\pi_i}$ denote the fraction of vertices $v \in S$ for which $R^G_{\pi_i}(v) > \ell$
	\If {$\tilde{p}_{\pi_i} < 3\gamma/4$}
		\State \textbf{return} $\pi = \pi_i$
	\EndIf
\EndFor
\State \textbf{report} ERROR
\EndProcedure
\end{algorithmic}
\end{algorithm}

Now we show that with probability at least $1-\delta$, Algorithm \ref{alg.MIS.1} finds a good $\pi$.
Let $H_{\pi, v}$ be the indicator variable such that $H_{\pi, v} = 1$ when $R^G_{\pi}(v) > \ell$, and $H_{\pi, v} = 0$ otherwise.
By Markov's inequality and Theorem~\ref{YYI} it holds that 
\BEQst
\Pr_{\pi \in S_n, v \in V} [R^G_{\pi}(v) > \ell] \leq \gamma/6\;.
\EEQst
That is, 
\BEQ
\E_{v \in V}[\E_{\pi \in S_n} [H_{\pi, v}]] \leq \gamma/6\;.\label{ves.eq}
\EEQ 
%Since we can determine whether $R^G_{\pi}(v) >\ell$ by querying $\pi$ on at most $\ell$ locations, it follows 
Next we aim to justify that we may obtain our random permutations by projecting orderings from $D_\ell$ instead of drawing directly from $S_n$.
More specifically, we claim that for a fixed $v$, $H_{\pi, v}$ is identically distributed regardless of whether $\pi$ is drawn uniformly from $S_n$ or obtained by projecting an ordering $r$ drawn from $D_\ell$.
To establish this claim, consider the evaluation of $H_{\pi, v}$ by a decision tree $\mathcal{T}_{G, v}$,
where each inner vertex of $\mathcal{T}_{G, v}$ represents a query to the rank $r(u)$ identified with the queried vertex $u\in V$. 
In order to evaluate $H_{\pi, v}$ we proceed down the tree $\mathcal{T}_{G, v}$ according to $\pi$ until we reach a leaf. 
After at most $\ell$ queries to $\pi$, the value of the indicator $H_{\pi, v}$ is determined. 
Therefore, the depth of $\mathcal{T}_{G, v}$ is at most $\ell$, and each leaf is associated with either $0$ or $1$. 
In other words, each leaf is identified with a sequence of at most $\ell$ vertices and their corresponding ranks. 

Let $s_1$ and $s_2$ be a pair of leaves such that:
(1) $s_1$ and $s_2$ are identified with the same sequence of vertices, $v_1, \ldots, v_k$; and  
(2) the ranks of $v_1, \ldots, v_k$ in $s_1$ and $s_2$ induce the same permutation on $v_1, \ldots, v_k$. 
Clearly the values corresponding to leaves $s_1$ and $s_2$ are identical. 
Therefore, our claim follows from the definition of $\ell$-wise independent ordering.
Thus, from equation (\ref{ves.eq}), we obtain
\BEQ
\E_{v \in V}[\E_{r \in D_\ell} [H_{\pi, v}]] \leq \gamma/6\;.\label{vve.eq}
\EEQ 

Next, we say that $\pi \in S_n$ is {\em very good} if $\Pr_{v\in V}\left[ R^G_{\pi}(v) > \ell\right] \leq \gamma/2$. 
By Markov's inequality and Equation~(\ref{vve.eq}), an ordering $r$ drawn from $D_\ell$ induces a very good permutation with probability at least $2/3$.
Therefore an ordering $r$ drawn according to a distribution that is $(1/n^2)$-almost $\ell$-wise independent random ordering over $[n]$ induces a very good permutation with probability at least $2/3 - 1/n^2$.
So by generating $\Theta(\log (1/\delta))$ permutations $\pi_i$'s from such random ordering, with probability greater than $1-\delta/2$, at least one of them must be very good.

Let $E_1$ denote the event that none of the selected $\pi_i$'s is very good, so $E_1$ occurs with probability smaller than $\delta/2$.
Let $E_2$ denote the event that there exists a selected $\pi_i$ such that $|p_{\pi_i} - \tilde{p}_{\pi_i}| > \gamma n/4$.
By Chernoff's bound and the union bound, $E_2$ occurs with probability at most $\delta/2$.
Given that $E_2$ does not occur and that $\tilde{p}_{\pi_i} \leq 3\gamma/4$, it follows that $\pi_i$ is good; i.e., $p_{\pi_i} \leq \gamma$. 
Therefore, given that both $E_1$ and $E_2$ do not occur, Algorithm \ref{alg.MIS.1} indeed finds a good ordering. 
By the union bound, we obtain a good ordering $\pi$ with probability at least $1-\delta$, as desired.

Theorem~\ref{alg.arv} implies that a random seed of length $O((\bar{d}/\eps)\log^2 n)$ suffices in order to obtain a $(1/n^2)$-almost $\ell$-wise independent random ordering over $[n]$.
Since Algorithm \ref{alg.MIS.1} draws $O(\log(1/\delta))$ such random orderings, overall a seed of length $O((\bar{d}/\eps)\log(1/\delta)\log^2 n)$ suffices. 
Additionally, we need $O((d^2/\eps^2)\log (1/\delta) \log \log (1/\delta) \log n)$ more random bits to determine the set $S$ of sampled vertices for evaluating $\tilde{p}_{\pi}$.
For Algorithm \ref{alg.MIS.1} to succeed with high probability, it suffices to substitute $\delta = 1/\poly(n)$, which yields the seed length claimed in the theorem statement. 

Finally, we now turn to describe the second phase of the algorithm.
Let $I_\pi$ denote the maximal independent set greedily created by $\pi \in S_n$, as defined by Algorithm~\ref{alg.MIS}.
Consider the following LCA $\mathcal{L}_\pi$ for a good permutation $\pi$.
On query $v$, $\mathcal{L}_\pi$ simulates the execution of $\textsc{LS-MIS}(\mathcal{O}^G, \pi, v)$ until it performs up to $\ell$ recursive calls to $\textsc{LS-MIS}$.
If $\textsc{LS-MIS}(\mathcal{O}^G, \pi, v)$ terminates within $\ell$ steps, $\mathcal{L}_\pi$ returns the answer given by $\textsc{LS-MIS}(\mathcal{O}^G, \pi, v)$.
Otherwise it simply returns NO.

Let $I'_\pi = I_\pi \cap V'_\pi$ where $V'_\pi = \{v : R^G_{\pi}(v) \leq \ell\}$, and notice that $\mathcal{L}_\pi$ answers YES precisely on $I'$.
Since $\pi$ is good, we have that $V'_\pi \geq (1-\gamma) n$ and so $|I'_\pi| \geq |I_\pi| -\gamma n$.
Since for any maximal independent set $I$ it holds that $|I| \geq n/d$, we obtain $|I'_\pi| \geq (1-\eps) |I_\pi|$.
In other words, $\mathcal{L}_\pi$ computes a $(1-\epsilon)$-approximate MIS, as required.
\EPF

\subsection{LCA for Computing an Approximate Maximum Matching}

Now we turn to describe an LCA that provides a $(1-\eps)$-approximation to maximum matching.
Our LCA locally simulates the global algorithm based on the following theorem by Hopcroft and Karp:

\BL
\cite{hopcroft1973n} Let $M$ and $M^*$ be a matching and a maximum matching in $G$. If the shortest augmenting path with respect to $M$ has length $2i-1$, then $|M|\leq (1-1/i)|M^*|$.
\EL

This global algorithm is used in the many contexts, including distributed computing and approximation (e.g., \cite{lotker2008improved,nguyen2008constant, yoshida2009improved}).
This algorithm can be summarized as follows. Let $M_0$ be an empty matching, and let $k = \lceil 1/\epsilon \rceil$.
We repeat the following process for each $i=1,\ldots, k$. Let $P_i$ denote the set of augmenting paths of length $2i-1$ with respect to $M_{i-1}$.
We compute a maximal set $A_i$ of vertex-disjoint augmenting paths, then augment these paths to $M_{i-1}$ to obtain $M_i$;
that is, we set $M_i = M_{i-1} \Delta A_i$ where $\Delta$ denotes the symmetric difference between two sets.
As a result, $M_k$ will be a $(1-\epsilon)$-approximate maximum matching.

The local simulation of this algorithm uses the formula $M_i = M_{i-1} \Delta A_i$ to determine whether the queried edge $e$ is in $M_k$ in a recursive fashion.
To determine $A_i$, consider the following observation.
Let $H_i = (P_i, E_i)$ be the graph whose vertex set corresponds to the augmenting paths, and $(u, v) \in E_i$ if and only if the augmenting paths $u$ and $v$ share some vertex.
$A_i$ is a maximal set of vertex-disjoint augmenting paths; that is, $A_i$ is an MIS of $H_i$.
In order to compute locally whether a path is in $A_i$ or not, Yoshida et al.~apply Algorithm \ref{alg.MIS} on $H_i$.
They show that in expectation over the queries and the random permutations of paths in $P_1, \ldots, P_k$, the query complexity is polynomial in $d$.
More formally, let $\vec{\pi} = (\pi^1, \ldots, \pi^k)$ where $\pi^i$ is a permutation of vertices (augmenting paths) in $P_i$, and let $Q^G_{\vec{\pi},k}(v)$ denote the query complexity of this local simulation.
Yoshida et al.~prove the following theorem.
\BT[\cite{yoshida2009improved}] \label{thm.mm2}
For any graph $G = (V, E)$ with $n$ vertices and maximum degree $d$, 
\BEQst
\E[Q^G_{\vec{\pi},k}(v)] \leq d^{6k^2}k^{O(k)}\;,\label{eq.mm}
\EEQst
where $\vec{\pi} = (\pi^1, \ldots, \pi^k)$ and the expectation is taken over the uniform distribution over $S_{|P_1|}\times \ldots \times S_{|P_k|}\times |E|$.
\ET

Based on this theorem, we convert the aforementioned local simulation into an LCA as similarly done in the previous section for the approximate MIS problem.
We now prove the following theorem.

\BT\label{thm.mm}
There is a randomized $(1-\eps)$-approximation local computation algorithm for maximum matching with
%random seed of length $O((d^{6k^2+1}k^{O(k)}/\eps) \log^3 n  \log \log n)$ and query complexity $O((d^{6k^2+2}k^{O(k)}/\eps^2) \log^2 n \log \log n)$ where $k = \Theta(1/\eps)$.
random seed of length $O(d^{6k^2+1}k^{O(k)} \log^3 n  \log \log n)$ and query complexity $O(d^{6k^2+2}k^{O(k)} \log^2 n \log \log n)$ where $k = \Theta(1/\eps)$.
\ET
\BPF
We build on the local algorithm of Yoshida et al.~(\cite{yoshida2009improved}), but here we double the number of iterations to $k =  \lceil 2/\epsilon \rceil$ so that $M_k$ is a $(1-\epsilon/2)$-approximate maximum matching.
The proof for this theorem follows the structure of the proof of Theorem~\ref{thm.mis}. 
Our LCA contains two phases: the first phase finds a sequence of good permutations, then the second phase locally simulates the approximate maximum matching algorithm using those permutations.
We say that a sequence of permutations $\vec{\pi} = (\pi_1, \ldots, \pi_k) \in S_{|V_1|}\times \ldots \times S_{|V_k|}$ is {\em good} if $\Pr_{e \in E}\left[ Q^G_{\vec{\pi}}(e) > \ell\right] \leq \gamma$ where $\gamma = \eps/(2d)$, $\ell = 6t/\gamma$ and $t = d^{6k^2}k^{O(k)}$.

In the first phase, rather than individually generating a permutation $\pi^k$ of $\vec{\pi}$ on each $P_i$, we create a single permutation $\pi$ over $[m_{n,k}]$ where $m_{n,k} =  \sum_{i=1}^k {n \choose k} \geq \sum_{i=1}^k |P_i|$.
In terms of constructing random orderings, this simply implies extending the domain of our orderings to cover all augmenting paths from all $k$ iterations.
Then for each $i\in [k]$, the permutation $\pi^k$ of paths $P_i$ can be obtained by restricting $\vec{\pi}$ to $P_i$ (i.e., considering the relative order among paths in $P_i$).
Since the algorithm queries $\vec{\pi}$ on at most $\ell$ locations, we conclude that Theorem \ref{thm.mm2} holds for any $\vec{\pi}$ that is $\ell$-wise independent.
Similarly to the proof of Theorem~\ref{thm.mm}, with probability at least $1-1/\poly(n)$, we shall find a good $\vec{\pi}$ by testing $\Theta(\log n)$ random orderings obtained via the construction of Alon et al.~in Theorem \ref{alg.arv}. 
Therefore, to this end we need at most $O(\ell \log^3 n +(d^2/\eps^2)\log^2 n \log \log n)$ random bits, as required.

Now that we obtain a good sequence of permutations $\vec{\pi}$ from the first phase, we again perform the altered version of local algorithm that returns NO as soon as the simulation invokes too many queries.
Let $M_{\vec{\pi}}$ denote the matching obtained by the algorithm of Yoshida et al.~when executed with the permutations from $\vec{\pi}$.
Let $M'_{\vec{\pi}}  = M_{\vec{\pi}} \cap E'_{\vec{\pi}}$ where $E'_{\vec{\pi}} = \{e : Q^G_{\vec{\pi}}(v) \leq \ell\}$. 
Similarly to the proof of Theorem~\ref{thm.mis}, given $\vec{\pi}$, we obtain an LCA that answers according to $M'_{\vec{\pi}}$ with query complexity $\ell$.
If $\vec{\pi}$ is good, we have that $E'_{\vec{\pi}} \geq (1-\gamma) |E|$ and so $|M'_{\vec{\pi}}| \geq |M_{\vec{\pi}}| -\gamma |E| \geq (1-\eps/2) |M^*| - \gamma |E|$ where $M^*$ denotes a maximum matching.
Since for any maximal matching $M$ it holds that $|M| \geq |E|/(2d)$ we obtain $|M'_{\vec{\pi}}| \geq (1-\eps) |M^*|$, as required.
\EPF
\section{Acknowledgments}
We thank Dana Ron for her valuable contribution to this paper.

\ifnum\icalp=0
\bibliographystyle{alpha}
\else
\bibliographystyle{plain}
\fi

\bibliography{bib}
\appendix

\section{Proof of Lemma \ref{v-active}} \label{lemproof}

\revactive*
\BPF
For each $u\in\Gamma_{G'}(v)$, let $E_u$ denote the event where $u$ is the only vertex in $\Gamma^+_{G'}(\{u, v\})$ that selects itself.
Since the maximum degree in $G'$ is at most $d/2^{j-1}$, then \[\Pr[E_u]=p_j (1-p_j)^{|\Gamma^+_{G'}(\{u, v\})|} \geq p_j (1-p_j)^{2d/2^{j-1}} \geq \frac{p_j}{e^2}.\]

Notice that $E_u$ is disjoint for each $u\in\Gamma_{G'}(v)$. Since $v\in V_j$, then $\deg_{G'}(v)\geq d/2^j$.
Thus $v$ becomes inactive with probability at least \[\sum_{u\in\Gamma_{G'}(v)}\Pr[E_u] \geq \deg_{G'}(v) \cdot \left(\frac{p_j}{e^2}\right) \geq \frac{1}{4e^2}.\]
That is, the theorem holds with parameter $p = 1 - 1/4e^2$.
\EPF

\section{Details of Theorem \ref{thm-mm2}} \label{thm-mm2-proof}

\resmmt*

In \cite{israeli1986fast}, Israeli and Itai proposed a randomized distributed algorithm which takes $O(\log n)$ rounds to compute a maximal matching.
Similarly to the MIS problem, Barenboim et al. also create a variant of this algorithm that, within $O(\log d)$ rounds,
finds a large matching that breaks the remaining graph into small connected components \cite{barenboim2012locality}.
Specifically, by running Algorithm \ref{dist-MM-phase1}, the remaining graph satisfies the following lemma.

\begin{algorithm*}
\caption{Barenboim et al.'s variant of Israeli and Itai's algorithm for computing a partial matching (simplified)}\label{dist-MM-phase1}
\begin{algorithmic}[1]
\Procedure{Distributed-MM-Phase1}{$G,d$}
	\State initialize matching $M = \emptyset$
	\For {$i = 1, \ldots, c_2 \log d$} \Comment $c_2$ is a sufficiently large constant
		\State initialize directed graphs $F_1 = (V, \emptyset)$ and $F_2 = (V, \emptyset)$
		\State {each vertex $s$ chooses a neighbor $t$ (if any) uniformly at random,
			\par\hspace{2.0em} then add $(s,t)$ to $E(F_1)$}
		\State {each vertex $t$ with positive in-degree in $F_1$ chooses a vertex $s \in \{s': (s',t)\in E(F_1)\}$ 
			\par\hspace{2.0em} with highest ID, then add $(s,t)$ to $E(F_2)$}
		\State each node $v$ with positive degree in $F_2$ chooses a bit $b(v)$ as follows:
			\par\hspace{2.0em} if $v$ has an outgoing edge but no incoming edge, $b(v) = 0$
			\par\hspace{2.0em} if $v$ has an ingoing edge but no outcoming edge, $b(v) = 1$
			\par\hspace{2.0em} otherwise, chooses $b(v)\in\{0,1\}$ uniformly at random
		\State add every edge $(s,t)\in E(F_2)$ such that $b(s)=0$ and $b(t)=1$ to $M$,
			\par\hspace{2.0em} and remove matched vertices from $G$
	\EndFor
\EndProcedure
\end{algorithmic}
\end{algorithm*}

\BL[\cite{barenboim2012locality}]\label{MM-lemma} 
\textsc{Distributed-MM-Phase1}$(G,d)$ computes a partial matching $M$ of an input graph $G$ within $O(\log⁡ d)$ communication rounds, such that the remaining graph contains no connected component of size larger than $O(d^4 \log ⁡n)$ with probability at least $1-1/\poly(n)$.
\EL

The proof of this lemma also makes use of Beck's analysis,
but contains a more complicated argument which shows that with constant probability,
each vertex loses some constant fraction of its neighbors in every round.
Thus, applying such matching subroutine for $O(\log d)$ rounds suffices to remove or isolate each vertex with probability $1-1/\poly(d)$.
We convert this lemma into a two-phase LCA in a similar fashion to obtain the desired LCA.

\end{document}